\documentclass{article}
\usepackage{spconf,amsmath,graphicx,hyperref}
\usepackage{amssymb,amsfonts}
\usepackage{algorithmic}
\usepackage{booktabs}
\usepackage{tabularx}
\usepackage{textcomp}
\usepackage{xcolor}
\usepackage{gensymb}
\usepackage{placeins}
\usepackage[T1]{fontenc}

\title{FUN-SSL: Full-band Layer Followed by U-Net with Narrow-band Layers for Multiple Moving Sound Source Localization}
\name{Yuseon Choi, Hyeonseung Kim, Jewoo Jun, Jong Won Shin \thanks{This work was partly supported by the Institute of Information \& Communications Technology Planning \& Evaluation (IITP) - ITRC (Information Technology Research Center) grant funded by the Korea government (MSIT) (IITP-2025-RS-2021-II211835, 50\%) and IITP grant funded by the Korea government (MSIT) (No.IRIS-2025-25443882, S.A.M.A.N.T.H.A: Sentient Audio Machine for Alive Natural Talking \& Human Affection, 50\%).}}
\address{Gwangju Institute of Science and Technology\\
\{newsun0130, kimhs355, zeusfront\}@gm.gist.ac.kr, jwshin@gist.ac.kr}
\begin{document}
%
\maketitle
\begin{abstract}

Dual-path processing along the temporal and spectral dimensions has shown to be effective in various speech processing applications. While the sound source localization (SSL) models utilizing dual-path processing such as the FN-SSL and IPDnet demonstrated impressive performances in localizing multiple moving sources, they require significant amount of computation. In this paper, we propose an architecture for SSL which introduces a U-Net to perform narrow-band processing in multiple resolutions to reduce computational complexity. The proposed model replaces the full-narrow network block in the IPDnet consisting of one full-band LSTM layer along the spectral dimension followed by one narrow-band LSTM layer along the temporal dimension with the FUN block composed of one Full-band layer followed by a U-net with Narrow-band layers in multiple scales. On top of the skip connections within each U-Net, we also introduce the skip connections between FUN blocks to enrich information. Experimental results showed that the proposed FUN-SSL outperformed previously proposed approaches with computational complexity much lower than that of the IPDnet. 
\end{abstract}
\section{Introduction}
Sound source localization (SSL) aims to estimate  the spatial position of one or multiple sound sources with respect to a given microphone array based on multi-channel audio signals. 
The output of the SSL is exploited in a variety of downstream applications such as sound source separation, speech enhancement, and automatic speech recognition \cite{multi_source_asr,survey, mcsess}.
Traditional approaches \cite{trad1, trad2, trad4, sapl1} apply statistical analyses on a certain spatial features to localize sound sources, but degrade in the presence of heavy noises and reverberations. 
Deep learning-based approaches \cite{sapl2, adap, robust, srsrp, doa, learning, uncertainty, fnssl, tfmamba} have shown superior performance in challenging acoustic scenarios when sufficient training data is available. Many of them, however, focus on localization of a single source or static sources, although the localization of multiple moving sources would widen the application of the SSL. 

To localize multiple moving sources in adverse environment, the SSL model needs to capture temporal context to track the moving sources and also exploit spectral correlation to deal with challenging scenarios. 
Commonly adopted network architectures for moving SSL include convolutional neural networks (CNN) \cite{exploit, framewise} and convolutional recurrent neural networks (CRNN) \cite{salsalite, seldnet, srpdnn}.
IPDnet \cite{ipdnet} employs dual-path processing along the temporal and spectral dimensions with the full-narrow (FN) network blocks, which achieves remarkable performance in localizing both single and multiple sources under static and dynamic conditions. The full-band layer along the spectral dimension captures inter-frequency correlations, whereas the narrow-band layer models temporal dynamics within individual frequency bands. While this structure was shown to be effective to track multiple moving sources, the computational complexity was rather high. 

In this study, we propose a modified network architecture to localize multiple moving sources more efficiently. 
As the U-Net \cite{unet, waveunet} can 
effectively capture multi-scale information across different resolutions with relatively low computational cost, we have adopted a U-net structure into the FN network blocks in the original IPDnet. 
In each repeated block, a \textbf{F}ull-band BLSTM layer is followed by a \textbf{U}-Net architecture equipped with \textbf{N}arrow-band LSTM layers in multiple scales, which we refer to as the FUN block.
By effectively integrating temporal features across multiple resolutions, FUN-SSL effectively estimates the direct-path relative transfer functions (DP-RTFs) of multiple moving sound sources. To provide richer information to the following FUN blocks, we introduce the skip connections between subsequent FUN blocks at the same resolution like the 
inter-U-Net skip connections in \cite{cunet} on top of the skip connections within each U-Net. 
Experimental results on a simulated dataset demonstrated that the proposed method, FUN-SSL, outperformed previously proposed methods with a comparable model size and reduced computational cost.
\section{Background}

\begin{figure*}[t]
    \centering
    \includegraphics[width=\textwidth]{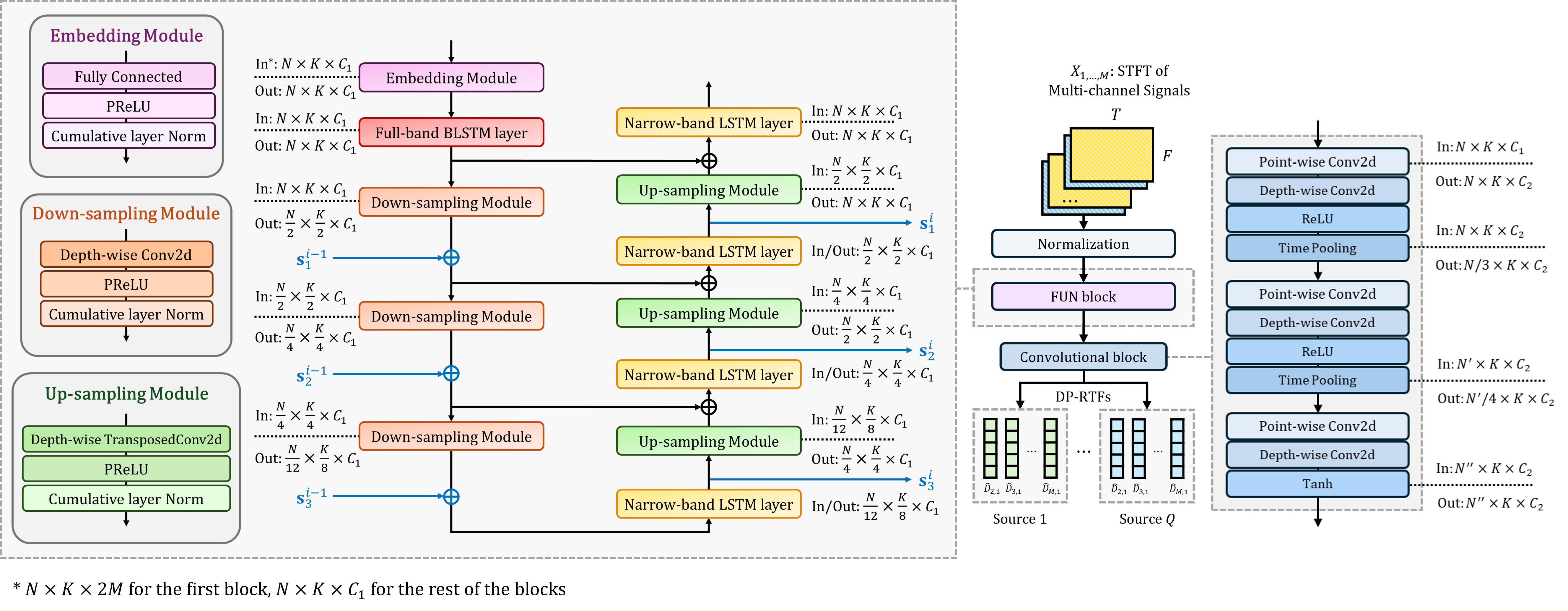} 
    \caption{Network Architecture of the proposed FUN-SSL.}
    \label{fig:overall}
\end{figure*}

Assuming a free- and far-field scenario, the signal captured at the $m$-th microphone in the short-time Fourier transform (STFT) domain for the $n$-th frame and the $k$-th frequency bin, $X_m(n, k)$, can be modeled as
\begin{equation}
    X_m(n, k) = \sum_{p=1}^{P} A_m(k, \theta_p(n)) S_p(n, k) + V_m(n, k),
\end{equation}
where 
$S_p(n,k)$ is the $p$-th source signal out of $P$ sources, 
$\theta_p(n)$ represents the direction of arrival (DoA) for $p$-th source at the $n$-th time frame, $A_m(k, \theta_p(n))$ is the direct-path acoustic transfer function (DP-ATF), and $V_m(n,k)$ is noise including all non-direct-path contributions.
The DP-RTF is defined as the ratio of the DP-ATFs for the $m$-th and the first microphones, which is given by 
\begin{align}
    \label{eq:dp-rtf}
    D_{m,1}(k, \theta_p(n))
    &= \frac{A_m(k, \theta_p(n))}{A_1(k, \theta_p(n))} 
       = \frac{e^{-j2\pi \nu_k \tau_m(\theta_p(n))}}
              {e^{-j2\pi \nu_k \tau_1(\theta_p(n))}} \\
    &= e^{-j2\pi \nu_k \Delta \tau_{m,1}(\theta_p(n))} \notag \\
    &= \cos(\mathrm{IPD}_{m,1}(k,\theta_p(n))) \notag \\
    &\quad + j\sin(\mathrm{IPD}_{m,1}(k,\theta_p(n))) \notag
\end{align}
in which $\nu_k$ is the frequency for the $k$-th bin, $\tau_m(\theta)$ is the time delay in the $m$-th microphone signal for the source coming from the direction $\theta$, $\Delta\tau_{m,1}(\theta)=\tau_m(\theta)-\tau_1(\theta)=\frac{d\cos(\theta)}{c}$ is the inter-channel time difference between the $m$-th and the first microphones where $d_m$ is the distance from the $m$-th microphone to the first one and $c$ is the speed of sound, and $\mathrm{IPD}_{m,1}(k,\theta)=-j2\pi\nu_k\Delta\tau_{m,1}(\theta)$ is the inter-channel phase difference (IPD) for the DoA $\theta_p(n)$.

IPDnet \cite{ipdnet} estimates DP-RTFs for multiple microphones using the full-band and narrow-band fusion network to localize multiple moving sources. In this paper, we propose a modified architecture which is computationally more efficient than the full-narrow network block of the IPDnet, while the input, output, training target and loss function stay the same. 
\section{Method}
\subsection{Overall architecture}\label{sec:overall}
The overall architecture of the proposed FUN-SSL is illustrated in Fig. \ref{fig:overall}.
The network receives the STFT representation of multi-channel audio signals $\mathbf{x} \in \mathbb{R}^{N\times K \times 2M}$ as input, where $N$, $K$, and $M$ are the numbers of frames, frequency bins, and microphones, and $2$ is for the real and imaginary parts. 
This input $\mathbf{x}$ is normalized by the Laplace normalization as in \cite{ipdnet}, and then processed by a series of FUN blocks. 
Each FUN block is composed of an embedding module, a full-band BLSTM layer, and a U-Net with narrow-band LSTM layers, which is composed of a sequence of down-sampling layers and a series of narrow-band LSTM layers followed by up-sampling layers.
The output from the last FUN block is finally processed by a causal convolutional block to produce the estimates for the real and imaginary parts of the DP-RTF vectors for each of the $M-1$ microphone pairs for all $Q$ sources \cite{ipdnet}, in which the depth-wise separable convolutions with channel dimension $C_{2}$ are applied instead of normal convolutions used in \cite{ipdnet} to reduce the number of parameters and computational complexity.
As in \cite{ipdnet}, the estimates for DP-RTFs are obtained once for 12 frames with the time pooling layers. 
The DP-RTF estimates for each of the $Q$ candidate sources are compared with the theoretical DP-RTFs for possible directions and then the source activity and the DoA are determined. 

\subsection{FUN block}\label{sec:FUN}
The FN block in IPDnet \cite{ipdnet} consists of a full-band BLSTM layer and a narrow-band LSTM layer with skip connections that concatenate the input of the block into the output of each layer to enrich the information. 
To reduce the computational complexity and exploit temporal correlations in multiple scales, we propose to employ FUN blocks which is a dual-path processing block built upon the FN block. 
FUN block consists of an embedding module, a full-band BLSTM layer, down-sampling and up-sampling modules, and narrow-band LSTM layers in multiple resolutions. FUN block does not have any skip connection from the block input which contributes to a significant amount of computational complexity, but has skip connections inside the U-Net and between the neighboring FUN blocks in multiple scales. 

Unlike the FN block in which the input is directly fed into a full-band BLSTM layer, FUN block introduces the embedding module composed of a fully-connected (FC) layer, a PReLU activation function, and cumulative layer normalization (cLN) \cite{convtasnet}. 
Since the subsequent down-sampling and up-sampling modules perform depth-wise convolutions only, a fully connected layer is employed to mix information across channels and compensate for this limitation.
For the $i$-th FUN block, the processing within the embedding module is represented as
\begin{equation}
    \mathbf{x}_{\mathrm{out}}^{i}=\mathrm{cLN}\left(\mathrm{PReLU}(\mathrm{FC}(\mathbf{x}_\mathrm{in}^{i}))\right)\in \mathbb{R}^{N\times K \times C_1},    
\end{equation}
where $\mathbf{x}_{\mathrm{in}}^{i}\in\mathbb{R}^{N\times K \times C_1}, i\geq 2$, and $\mathbf{x}_{\mathrm{in}}^{0}=\mathbf{x}\in\mathbb{R}^{N\times K \times 2M}$.

The output of the embedding module is then processed by a single full-band BLSTM layer as in the FN block, i.e., 
\begin{equation}
\mathbf{d}_{0}^{i}=\mathrm{BLSTM}_\mathrm{full}(\mathbf{x}_\mathrm{out}^{i})\in \mathbb{R}^{N\times K \times C_1},
\end{equation}
where $\mathbf{d}_{0}^{i}$ denotes the features after full-band processing, and the layer $\mathrm{BLSTM}_{\mathrm{full}}$ operates along the frequency axis.

The output of the full-band BLSTM layer is processed by a U-Net with narrow-band LSTM layers detailed in Fig. \ref{fig:overall}. 
Each of three down-sampling modules comprises a depth-wise 2D convolutional layer, PReLU, and cLN. 
As in \cite{sudormrf}, we employ a depth-wise convolution instead of a standard convolution for more efficient processing. 
The output from the $j$-th down-sampling module becomes 
\begin{equation}
\mathbf{d}_{j}^{i}
= \mathrm{cLN}\left(
    \mathrm{PReLU}\left(
        \mathrm{DConv}_{(5, 2h_j)}^{(2, h_j)}\left(
            \mathbf{d}^{i}_{j-1}
        \right)
    \right)
\right) + \mathbf{s}_{j}^{i-1},
\end{equation}
in which $h_j$ is the down-sampling factor set to be $h_1=2$, $h_2=2$, and $h_3=3$, and the kernel size and stride for the depth-wise convolution is $(5,\ 2h_j)$ and of $(2,\ h_j)$. $\mathbf{s}_j^{i-1}$ is the output of the narrow-band LSTM layer operating at the same scale with $\mathbf{d}_{j}^{i}$ in the $(i-1)$-th FUN block connected with a inter-U-Net skip connection. 
For the first block, $\mathbf{s}_{j}^{0}$ is initialized as a zero vector.

For each scale, the signal is processed by a narrow-band LSTM layer. The output of the $j$-th narrow-band LSTM layer is denoted as $\mathbf{s}^i_{4-j}$ as it has the same scale with the $(4-j)$-th down-sampling module output, and is given by 
\begin{equation}
\label{eq:nlstm}
\mathbf{s}^i_{4-j} 
= \mathrm{LSTM}_\mathrm{narrow}(\mathbf{u}^i_{j-1} + \mathbf{d}_{4-j}^{i} ), ~~j=1, 2, 3, 4, 
\end{equation}
in which $\mathbf{u}^i_{j-1}$ is the output of the $(j-1)$-th up-sampling module with $\mathbf{u}^i_{0}=\mathbf{0}$. Here, $\mathrm{LSTM_{narrow}}$ operates along the time axis.
Each of three up-sampling modules takes $\mathbf{s}^i_{4-j}$ as input and processes it with a depth-wise transposed 2D convolutional layer, PReLU, and cLN, i.e., 
\begin{equation}
\label{eq:usample}
\mathbf{u}^i_j 
= \mathrm{cLN}\left( 
    \mathrm{PReLU}\left(
        \mathrm{DConvTransposed}_{(5, h_{4-j})}^{(2, h_{4-j})}\left(
            \mathbf{s}^i_{4-j}
        \right)
    \right)
\right).
\end{equation}
The output of the last narrow-band LSTM layer, $\mathbf{s}^i_{0}$, becomes the output of the $i$-th FUN block.

It is noted that in the causal convolutional block which processes the output of the last FUN block with a depth-wise separable convolutions, 
the point-wise convolution is applied prior to the depth-wise convolution unlike the conventional depth-wise separable convolution, as the up-sampling modules in the FUN block consist of depth-wise convolutions. 


\section{Experimental Setup}
The proposed model was trained and evaluated on a simulated dataset with a sampling rate of 16 kHz similar to the one used in \cite{srpdnn} and \cite{ipdnet}. 
The microphone signals were synthesized by convolving clean speech signals from the LibriSpeech \cite{libri} corpus with room impulse responses (RIRs) generated by gpuRIR \cite{gpurir}. Reverberation time (RT60) was sampled between 0.2 s and 1.3 s, and room dimensions were randomly selected within the range of 6×6×2.5~m to 10×8×6~m. A maximum of two static or moving sound sources were considered. 
The proportions for static and moving sources as well as those for a single source and two sources cases were 50\% and 50\%. 
As in \cite{robust, srpdnn, fnssl, ipdnet}, the moving trajectory of each sound source was constructed by selecting random start and end points within the room, connecting them with a straight line, and adding sinusoidal perturbations along each axis (x, y, z) to generate curved paths. Two microphones were placed at random positions on the same horizontal plane with an inter-microphone distance of 8 cm. Diffuse noises generated by the ANF generator \cite{ANFgen} from the white, babble, and factory noise from the NOISEX-92 database \cite{noisex} were added at a random signal-to-noise ratio (SNR) between -5 dB and 15 dB. 300,000, 4,000, and 4,000 samples with a duration of 4.5 s were generated to form the training, validation, and test sets, respectively. 
SRP-DNN \cite{srpdnn} and IPDnet \cite{ipdnet} were compared.

The length of a window was 512 samples with a 50\% overlap, and 512 point STFT was applied. 
The maximum number of sources $Q$ was set to $2$. 
The threshold for sound source activity detection was set so that the miss detection rate (MDR) and the false alarm rate (FAR) became the same. 
The number of FUN blocks was set to 2. The channel dimensions for (B)LSTM layer and convolutional layer were configured as $C_1=96$ and $C_2=128$, respectively. As in the IPDnet \cite{ipdnet}, a permutation-invariant training (PIT) was employed with a mean squared error loss between the target and estimated DP-RTF. The Adam optimizer was used and the batch size was 16. The model was optimized for 40 epochs using a learning rate initialized to 0.001 with an exponential decay factor of 0.95.


The performance of source localization was evaluated only for speech-active frames. The azimuth candidates were discretized at a resolution of 1$\degree$, and the angular estimation error was defined as the absolute difference between the estimated and the target azimuth. Gross accuracy measures the ratio of the speech active frames detected and the error is less than $\mathrm{ET}=10\degree$, FAR is the number of estimated sources which are not active or the error are more than $\mathrm{ET}$ divided by the number of active sources, and Fine Error is the mean absolute error for the frames with the errors less than $\mathrm{ET}$ \cite{ipdnet, metric}. 

\section{Results}
The performance and computational complexity of each localization model is presented in Table \ref{tab:1}. It is noted that the performances for SRP-DNN \cite{srpdnn} and IPDnet \cite{ipdnet} are from the corresponding papers, for which the test data were constructed in a similar manner but not identical to the test data used to evaluate FUN-SSL. The experimental results using the official code from the authors\footnote{\url{https://github.com/Audio-WestlakeU/FN-SSL}} are denoted as IPDnet$^\dag$. The performance of the SRP-DNN \cite{srpdnn} was inferior to other models, although it was a much lighter model. The IPDnet$^\dag$ showed slightly better performance to that reported in the paper, and FUN-SSL outperformed IPDnet$^\dag$ in all metrics with slightly more parameters and almost half of the computations. 
\begin{table}[tb]
\centering
\caption{Complexity and performance comparison with previously proposed methods. $^\dag$: experiments with the code from the authors.}
\vspace{6pt} 
\resizebox{\linewidth}{!}{%
\begin{tabular}{@{}lccccc@{}}
\toprule
Model & \# Params. & FLOPs & Gross Acc & Fine Error & FAR \\ \midrule
SRP-DNN \cite{srpdnn}     & 0.8 M & 2.3 G/s  & 80.1 \% & 2.9\degree   & 13.1 \% \\
IPDnet \cite{ipdnet}      & 0.7 M & 19.4 G/s & 91.7 \% & 2.1\degree   & 7.7 \%  \\
IPDnet$^\dag$              & 0.7 M & 19.4 G/s & 93.0 \% & 2.0\degree   & 7.1 \%  \\
FUN-SSL                   & 0.8 M & 10.8 G/s & \textbf{94.2} \% & \textbf{1.9}\degree   & \textbf{5.8} \% \\
\bottomrule
\end{tabular}%
}
\label{tab:1}
\end{table}
Table \ref{tab:4} presents the ablation study for the number of FUN blocks in the FUN-SSL. The results implied that the second block was necessary for good performance, but the third block had a marginal impact on the performance. 

\begin{table}[tb]
\centering
\caption{Complexity and performance according to the number of FUN blocks. }
\vspace{6pt} 
\resizebox{\linewidth}{!}{%
\begin{tabular}{@{}cccccc@{}}
\toprule
{\# Blocks} & \# Params. & FLOPs {[}G/s{]} & Gross Acc & Fine Error & FAR \\ \midrule
1          & 0.4 M     & 5.6 G/s      & 91.9 \%              & 2.2\degree              & 8.1 \% \\
2          & 0.8 M     & 10.8 G/s     & 94.2 \%              & \textbf{1.9}\degree     & 5.8 \% \\
3          & 1.2 M     & 16.0 G/s     & \textbf{94.5} \%     & \textbf{1.9}\degree     & \textbf{5.9} \%
     \\ \bottomrule
\end{tabular}%
}
\label{tab:4}
\end{table}

Due to the multi-scale structure in the FUN block, it contains five (B)LSTM layers in total, while the FN block in the IPDnet has two (B)LSTM layers. To examine whether the performance improvement was from the increased number of LSTM layers or the proposed structure of the FUN block, we performed an additional experiment comparing the FUN-SSL using two FUN blocks with $C_1=96$ with the SSL replacing two FUN blocks with five FN blocks having 10 (B)LSTM layers in total, for which $C_1$ was adjusted to $56$ to have similar computational cost with the FUN-SSL. 
The results are shown in Table \ref{tab:5}. FUN-SSL achieved superior localization performance than the SSL with five FN blocks with the same compuational complexity. 
This confirmed that the performance improvement was not merely due to the increased number of LSTM layers, but stemmed from the effectiveness of the proposed FUN block architecture including U-Net architecture and inter-block skip connections. 

\begin{table}[tb]
\centering
\caption{Performance comparison with the SSL using FN blocks with the same number of LSTM layers and FLOPs.}
\vspace{6pt}
\resizebox{\linewidth}{!}{
\begin{tabular}{lccccccc}
\toprule
Block & \# Params. & FLOPs & Gross ACC & Fine Error & FAR \\
\midrule
2 FUN Blocks     & 0.8 M & 10.8 G/s & \textbf{94.2} \% & \textbf{1.9}\degree & \textbf{5.8} \% \\
5 FN Blocks    & 0.4 M & 10.8 G/s & 93.0 \%          & 2.2\degree          & 7.0 \% \\
\bottomrule
\end{tabular}
}
\label{tab:5}
\end{table}

\section{Conclusion}
In this paper, we propose a novel architecture for sound source localization improving the one for the IPDnet, FUN-SSL, which is composed of a full-band layer followed by a U-Net with multi-scale narrow-band layers. By introducing a U-Net structure with depth-wise convolutions, the temporal dependencies were exploited at multiple scales computationally efficiently. 
In addition, inter-U-Net skip connections are introduced to further enhance the model capacity by preserving important spatial information from previous blocks and propagating it to a subsequent processing stage. FUN-SSL outperformed the IPDnet with a comparable number of parameters and significantly lower computational cost, demonstrating the effectiveness of the proposed architecture.
\vfill\pagebreak

\ninept
\bibliographystyle{IEEEbib}
\bibliography{strings,refs}\label{sec:refs}

\begin{thebibliography}{10}

\bibitem{multi_source_asr}
A.~S. Subramanian, C.~Weng, S.~Watanabe, M.~Yu, and D.~Yu,
\newblock ``Deep learning based multi-source localization with source splitting and its effectiveness in multi-talker speech recognition,''
\newblock {\em Comput. Speech \& Lang.}, vol. 75, pp. 101360, 2022.

\bibitem{survey}
P.-A. Grumiaux, S.~Kitić, L.~Girin, and A.~Guérin,
\newblock ``A survey of sound source localization with deep learning methods,''
\newblock {\em J. Acoustical Soc. America}, vol. 152, no. 1, pp. 107, July 2022.

\bibitem{mcsess}
S.~Gannot, E.~Vincent, S.~Markovich-Golan, and A.~Ozerov,
\newblock ``A consolidated perspective on multimicrophone speech enhancement and source separation,''
\newblock {\em IEEE/ACM Trans. Audio, Speech, Lang. Process.}, vol. 25, no. 4, pp. 692--730, 2017.

\bibitem{trad1}
C.~Knapp and G.~Carter,
\newblock ``The generalized correlation method for estimation of time delay,''
\newblock {\em IEEE Trans. Acoust. Speech Signal Process.}, vol. 24, no. 4, pp. 320--327, 2003.

\bibitem{trad2}
R.~Schmidt,
\newblock ``Multiple emitter location and signal parameter estimation,''
\newblock {\em IEEE Trans. Antennas and Propag.}, vol. 34, no. 3, pp. 276--280, 1986.

\bibitem{trad4}
M.~Raspaud, H.~Viste, and G.~Evangelista,
\newblock ``Binaural source localization by joint estimation of ild and itd,''
\newblock {\em IEEE Trans. Audio, Speech, Lang. Process.}, vol. 18, no. 1, pp. 68--77, 2010.

\bibitem{sapl1}
H.~Song and J.~W. Shin,
\newblock ``Multiple sound source localization based on interchannel phase differences in all frequencies with spectral masks.,''
\newblock in {\em Interspeech}, 2021, pp. 671--675.

\bibitem{sapl2}
J.~Pak and J.~W. Shin,
\newblock ``Sound localization based on phase difference enhancement using deep neural networks,''
\newblock {\em IEEE/ACM Trans. Audio, Speech, Lang. Process.}, vol. 27, no. 8, pp. 1335--1345, 2019.

\bibitem{adap}
W.~He, P.~Motlicek, and J.~M. Odobez,
\newblock ``Neural network adaptation and data augmentation for multi-speaker direction-of-arrival estimation,''
\newblock {\em IEEE/ACM Trans. Audio, Speech, Lang. Process.}, vol. 29, pp. 1303--1317, 2021.

\bibitem{robust}
D.~Diaz-Guerra, A.~Miguel, and J.~R. Beltran,
\newblock ``Robust sound source tracking using srp-phat and 3d convolutional neural networks,''
\newblock {\em IEEE/ACM Trans. Audio, Speech, Lang. Process.}, vol. 29, pp. 300--311, 2020.

\bibitem{srsrp}
J.~H. Cho and J.~H. Chang,
\newblock ``Sr-srp: Super-resolution based srp-phat for sound source localization and tracking,''
\newblock in {\em Interspeech}, 2023, pp. 3769--3773.

\bibitem{doa}
D.~Diaz-Guerra, A.~Miguel, and J.~R. Beltran,
\newblock ``Direction of arrival estimation of sound sources using icosahedral cnns,''
\newblock {\em IEEE/ACM Trans. Audio, Speech, Lang. Process.}, vol. 31, pp. 313--321, 2022.

\bibitem{learning}
B.~Yang, H.~Liu, and X.~Li,
\newblock ``Learning deep direct-path relative transfer function for binaural sound source localization,''
\newblock {\em IEEE/ACM Trans. Audio, Speech, Lang. Process.}, vol. 29, pp. 3491--3503, 2021.

\bibitem{uncertainty}
R.~Pi and X.~Yu,
\newblock ``Uncertainty estimation for sound source localization with deep learning,''
\newblock {\em IEEE Trans. Instrumentation and Measurement}, 2024.

\bibitem{fnssl}
Y.~Wang, B.~Yang, and X.~Li,
\newblock ``Fn-ssl: Full-band and narrow-band fusion for sound source localization,''
\newblock in {\em Interspeech}, 2023, p. 3779–3783.

\bibitem{tfmamba}
Y.~Xiao and R.~K. Das,
\newblock ``Tf-mamba: A time-frequency network for sound source localization,''
\newblock in {\em Interspeech}, 2025, pp. 948--952.

\bibitem{exploit}
A.~Bohlender, A.~Spriet, W.~Tirry, and N.~Madhu,
\newblock ``Exploiting temporal context in cnn based multisource doa estimation,''
\newblock {\em IEEE/ACM Trans. Audio, Speech, Lang. Process.}, vol. 29, pp. 1594--1608, 2021.

\bibitem{framewise}
L.~Wang, Z.~Jiao, Q.~Zhao, J.~Zhu, and Y.~Fu,
\newblock ``Framewise multiple sound source localization and counting using binaural spatial audio signals,''
\newblock in {\em ICASSP}, 2023, pp. 1--5.

\bibitem{salsalite}
T.~N.~T. Nguyen, D.~L. Jones, K.~N. Watcharasupat, H.~Phan, and W.~S. Gan,
\newblock ``Salsa-lite: A fast and effective feature for polyphonic sound event localization and detection with microphone arrays,''
\newblock in {\em ICASSP}, 2022, pp. 716--720.

\bibitem{seldnet}
S.~Adavanne, A.~Politis, J.~Nikunen, and T.~Virtanen,
\newblock ``Sound event localization and detection of overlapping sources using convolutional recurrent neural networks,''
\newblock {\em IEEE Journal Select. Topics Sig. Process.}, vol. 13, no. 1, pp. 34--48, 2018.

\bibitem{srpdnn}
B.~Yang, H.~Liu, and X.~Li,
\newblock ``Srp-dnn: Learning direct-path phase difference for multiple moving sound source localization,''
\newblock in {\em ICASSP}, 2022, pp. 721--725.

\bibitem{ipdnet}
Y.~Wang, B.~Yang, and X.~Li,
\newblock ``Ipdnet: A universal direct-path ipd estimation network for sound source localization,''
\newblock {\em IEEE/ACM Trans. Audio, Speech, Lang. Process.}, 2024.

\bibitem{unet}
O.~Ronneberger, P.~Fischer, and T.~Brox,
\newblock ``U-net: Convolutional networks for biomedical image segmentation,''
\newblock in {\em MICCAI}, 2015, pp. 234--241.

\bibitem{waveunet}
D.~Stoller, S.~Ewert, and S.~Dixon,
\newblock ``Wave-u-net: A multi-scale neural network for end-to-end audio source separation,''
\newblock {\em arXiv:1806.03185}, 2018.

\bibitem{cunet}
A.~A. Albishri, S.~J.~H. Shah, and Y.~Lee,
\newblock ``Cu-net: Cascaded u-net model for automated liver and lesion segmentation and summarization,''
\newblock in {\em BIBM}, 2019, pp. 1416--1423.

\bibitem{convtasnet}
Y.~Luo and N.~Mesgarani,
\newblock ``Conv-tasnet: Surpassing ideal time--frequency magnitude masking for speech separation,''
\newblock {\em IEEE/ACM Trans. Audio, Speech, Lang. Process.}, vol. 27, no. 8, pp. 1256--1266, 2019.

\bibitem{sudormrf}
E.~Tzinis, Z.~Wang, and P.~Smaragdis,
\newblock ``Sudo rm-rf: Efficient networks for universal audio source separation,''
\newblock in {\em MLSP}, 2020, pp. 1--6.

\bibitem{libri}
V.~Panayotov, G.~Chen, D.~Povey, and S.~Khudanpur,
\newblock ``Librispeech: an asr corpus based on public domain audio books,''
\newblock in {\em ICASSP}, 2015, pp. 5206--5210.

\bibitem{gpurir}
D.~Diaz-Guerra, A.~Miguel, and J.~R. Beltran,
\newblock ``gpurir: A python library for room impulse response simulation with gpu acceleration,''
\newblock {\em Multimedia Tools and Applications}, vol. 80, no. 4, pp. 5653--5671, 2021.

\bibitem{ANFgen}
E.~A.~P. Habets, I.~Cohen, and S.~Gannot,
\newblock ``Generating nonstationary multisensor signals under a spatial coherence constraint,''
\newblock {\em J. Acoustical Soc. America}, vol. 124, no. 5, pp. 2911--2917, 2008.

\bibitem{noisex}
A.~Varga and H.~J.~M. Steeneken,
\newblock ``Assessment for automatic speech recognition: Ii. noisex-92: A database and an experiment to study the effect of additive noise on speech recognition systems,''
\newblock {\em Speech communication}, vol. 12, no. 3, pp. 247--251, 1993.

\bibitem{metric}
J.~Woodruff and D.~Wang,
\newblock ``Binaural localization of multiple sources in reverberant and noisy environments,''
\newblock {\em IEEE Trans. Audio, Speech, Lang. Process.}, vol. 20, no. 5, pp. 1503--1512, 2012.

\end{thebibliography}

\end{document}